\documentclass[a4paper, 12pt]{scrartcl}

\usepackage{caption}
\setkomafont{title}{\sffamily\bfseries}
\addtokomafont{subsection}{\normalsize}

\deffootnote{1em}{1em}{\makebox[1em][l]{\thefootnotemark}}

\usepackage[whole]{bxcjkjatype}
\usepackage[T1]{fontenc}

\usepackage{amsthm}
\newtheorem{theorem}{Theorem}
\newtheorem{lemma}[theorem]{Lemma}
\newtheorem{conjecture}[theorem]{Conjecture}

\usepackage[lining]{ebgaramond}
\usepackage[vvarbb]{newtxmath}
\usepackage{ebgaramond-maths}
\usepackage[defaultsans]{opensans}

\usepackage{mathrsfs}
\usepackage{graphicx}
\usepackage{hyperref}

\newcommand{\Nset}{\mathbb N}
\newcommand{\Zset}{\mathbb Z}

\newcommand{\density}{\mathrm D}
\newcommand{\classP}{\mathsf{P}}
\newcommand{\classNP}{\mathsf{NP}}
\newcommand{\classPSPACE}{\mathsf{PSPACE}}

\title{Proof of the Density Threshold Conjecture for Pinwheel Scheduling\thanks{This is the author's accepted version of the paper published in the \emph{Proceedings of the National Academy of Sciences} (PNAS), 123(32), e2530214123, 2026.  Theorem numbers correspond to those in the published version, though section and reference numbers differ.  Preliminary announcements of this work appeared at IPSJ SIG on Algorithms~\cite{Kaw24a, Kaw24b} and the 56th Annual ACM Symposium on Theory of Computing (STOC)~\cite{Kaw24}.}}

\author{Akitoshi Kawamura}

\date{\vspace{-1.3\baselineskip}}

\begin{document}

\maketitle

\begin{abstract}
In the pinwheel scheduling problem, each task $i$ is associated with a positive integer $a_i$ called its period, and we want to (perpetually) schedule one task per day so that each task $i$ is performed at least once every $a_i$ days.  An obvious necessary condition for schedulability is that the density, defined as the sum of execution rates $1/a_i$, does not exceed $1$.  We prove that all instances with density not exceeding $5 / 6$ are schedulable, as was conjectured by Chan and Chin in 1993.  Like some of the known partial progress towards the conjecture, our proof involves computer search for schedules for a large but finite set of instances.  A key idea in our reduction to these finite cases is to generalize the problem to fractional (non-integer) periods in an appropriate way.  As byproducts of our ideas, we obtain a simple proof that every instance with two distinct periods and density at most $1$ is schedulable, as well as a fast algorithm for the bamboo garden trimming problem with approximation ratio $4 / 3$.
\end{abstract}

\section{Introduction}

\begin{figure}
\centering 
\raisebox{-36.5pt}{\includegraphics[scale=0.66]{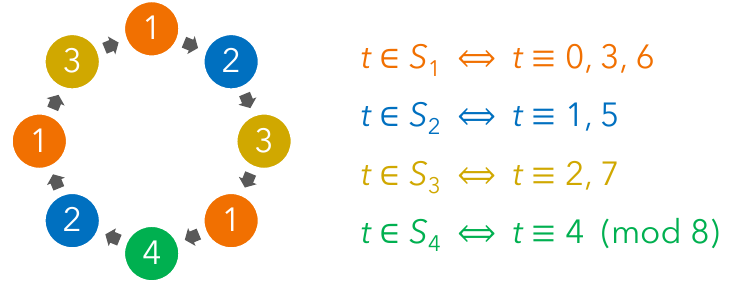}}
\caption{The instance $(3, 4, 5, 8)$ means that we have tasks $1$, $2$, $3$ and $4$ that
need to be performed at least once every $3$ days, $4$ days, $5$ days and $8$ days.
This is satisfied by the schedule $(S _1, S _2, S _3, S _4)$ that arranges the tasks in the $8$-day cycle as shown.}
\label{figure: schedule example}
\end{figure}

In many systems---%
like factory machines, sensor networks, or digital communication---%
tasks must be performed regularly, without long gaps. 
The \emph{pinwheel scheduling} (PS) problem \cite{HMRTV89} models this need: 
each task has a time limit within which it must be repeated, 
and the goal is to build a schedule that 
performs one task per unit time (or ``day'' herein)
while satisfying all these limits. 
Specifically,
each task $i \in [k] = \{1, \dots, k\}$
must be scheduled at least once every $a _i$
days, where $a _i$ is the \emph{period} of task $i$. 
Thus, an instance of PS
is an array $(a _i) _{i \in [k]} = (a _1, \dots, a _k)$ of positive integers,
which we assume to be arranged in non-decreasing order,
and we seek to find 
pairwise disjoint sets $S _1$, \ldots, $S _k \subseteq \Zset$ (with $S _i$ specifying the days on which to perform task $i$)
satisfying, for all $i \in [k]$, the 
\emph{frequency condition}:
\begin{quote}
for each $m \in \Zset$, 
we have $[m, m + a _i) \cap S _i \neq \emptyset$.
\end{quote}
\parshape=0
An instance for which such a 
\emph{schedule} $(S _i) _{i \in [k]}$ exists is said to be 
\emph{schedulable}.
For example, the instances
$(3, 3, 3)$, $(2, 4, 8, 8)$, and $(3, 4, 5, 8)$ are
schedulable (see Figure~\ref{figure: schedule example} for the last one), but
all become unschedulable if
any one period $a _i$ is replaced by $a _i - 1$.

For an instance $A = (a _i) _{i \in [k]}$ to be schedulable, the condition that its \emph{density}
\begin{equation*}
  \density (A) =
  \sum _{i \in [k]} \frac{1}{a _i}
\end{equation*}
does not exceed $1$ is clearly necessary, but not sufficient.
For example, $(2, 3, a _3)$ is unschedulable for
all values of $a _3$.
On the other hand, it is relatively easy to see that
every instance with a density of at most $\frac 1 2 = 0.5$
is schedulable~\cite[Corollary 3.2]{HMRTV89}.
The challenge of improving this sufficient condition has been taken up by several authors,
who succeeded in increasing the bound
to $0.66\ldots$~\cite{CC93},
to $0.7$~\cite[Theorem 4.2]{CC92},
and then to $0.75$~\cite[Theorem 1]{FL02}.
It has been conjectured~\cite{CC93}
that this value could be increased to $\frac 5 6 = 0.83\ldots {}$, 
which is the best possible
because of the instance $(2, 3, a _3)$ 
mentioned above.
This conjecture has been confirmed in
a number of special cases:
%\begin{itemize}
%\item
when $A$ has three (or fewer) distinct period values \cite[Theorem 4]{LL97},
%\item
when the smallest period~$a _1$ is $2$ \cite[Theorem 2]{FL02},
%\item
when for each period value there are at least five tasks having that period \cite[Theorem 3]{BL98},
%\item 
and when the number~$k$ of tasks is $\leq 12$ \cite{GSW22}.
%\end{itemize}
We resolve this conjecture affirmatively: 
\begin{theorem}
  \label{theorem}
Every PS instance that 
consists of positive integers and
has density $\leq \frac 5 6$ 
is schedulable.
\end{theorem}
The reason for the explicit restriction to integers
is that, in the following section,
we extend the PS problem and allow periods to be positive real numbers, not just integers. 
This extension plays a key role in our proof of Theorem~\ref{theorem}.

Following the preliminary announcements of this paper \cite{Kaw24a, Kaw24}, 
Fujiwara, Miyagi and Ouchi \cite{FMO26} conjectured, 
and confirmed for $k \leq 3$, 
that in fact Theorem~\ref{theorem} holds for 
possibly non-integer instances as well. 

\subsection*{Further Results}
The condition $\density (A) \leq 1$,
which is obviously necessary for any instance $A$,
has been shown also to be sufficient for instances~$A$
containing only two distinct periods \cite[Corollary 4.9]{HRTV92}. 
Our idea of fractional periods mentioned above
yields a simple proof of this statement 
(for possibly non-integer periods as well).
That is, we (re)prove: 

\begin{theorem}
\label{theorem: two distinct}
Every PS instance that consists of at most two distinct period values
and has density $\leq 1$ 
is schedulable.
\end{theorem}

\emph{Bamboo garden trimming} (BGT) \cite{GJKLLMR24}
is an optimization version of PS:
in a grove of $k$ bamboo plants,
plant $i \in [k]$ grows in height at a daily rate $h _i \in \Nset$;
each day (at a fixed time of day) we select
one plant to be trimmed (i.e., reduced to height $0$)
with the goal of keeping the highest plant in the grove as low as possible.
Clearly,
a trimming schedule achieves height $G$
(i.e., ensures that no plant ever grows beyond it) for this instance
if and only if
it satisfies the PS instance
$(\lfloor G / h _i \rfloor)_{i \in [k]}$.
Note that the smallest such $G$ is always an integer.  
Efficient algorithms to estimate this optimal $G$
have been proposed, with approximation ratios
$2$~\cite{GKLLMR17} (see also \cite{Kus22} for the discussion of how simple algorithms achieve or do not achieve this ratio),
$1.88\ldots$~\cite[Corollary 1]{DC20}, 
$1.71\ldots$~\cite[Corollary 1]{vE21},
$1.60\ldots$~\cite[Theorem 3]{GJKLLMR24},
and 
$1.42\ldots$~\cite[Section 3]{HvS23}.
We use Theorem~\ref{theorem},
and ideas used for it,
to obtain an improved approximation ratio $\frac 4 3 = 1.33\ldots {}$: 

\begin{theorem}
\label{theorem: bamboo}
There is a polynomial-time algorithm that, 
given a BGT instance, 
outputs a number $G \in \Nset$ such that 
the instance has a schedule with height $\leq \frac 4 3 \cdot G$
but none with height $< G$. 
\end{theorem}

Although Theorem~\ref{theorem: bamboo} is stated in terms of  
finding the optimal value approximately, 
it will be easy to see that our algorithm can be turned into one that finds an underlying schedule. 
But it is not clear whether it does so in ``polynomial time,''
because our schedule is an infinite object.
See the comments on ``fast online schedulers'' in Section~\ref{section: implementation}. 

After the preliminary announcements of Theorem~\ref{theorem: bamboo} were made \cite{Kaw24b, Kaw24}, 
Mishra \cite{Mis26} published further improvement on the approximation ratio. 

\subsection*{Related Work}
\label{subsection: related work}

The class of problems including ours and its variants below, 
where we seek perpetual schedules of
recurring tasks (such as monitoring, maintenance, or replenishment) 
that should be performed with sufficient frequencies, 
has been also called \emph{windows scheduling} or \emph{periodic scheduling}. 
These problems have found applications such as 
packet scheduling in communication networks \cite{HMRTV89,BL03}, 
machine maintenance \cite{WL83, GJKLLMR24}, 
security games \cite{KST18}, 
broadcasting commercials \cite{SST09},
and spaced repetition of review material in education \cite{NKS12}.

Our PS problem is perhaps the most basic among such settings. 
Some generalizations immediately present themselves:
one might suppose that multiple tasks can be performed each day~\cite{BL03, BLT07, MMNS26}, 
that different tasks take different amounts of time to perform~\cite{FC05, SST09, Kus26}, 
or that the required spacing between tasks changes over time~\cite{NKS12}. 
There are also variants with continuous flavour, 
where the tasks are placed on metric spaces
so that moving between them takes some time or cost 
\cite{CSW11, GJKLLMR24}, 
or even those problems, sometimes called \emph{patrolling}, 
where we look for continuous movements of agents 
so that all points in the terrain are visited with sufficient frequency
\cite{CGKK11, KK15, HKMPP19, KS20}. 
There are also various ways, besides BGT, to turn PS settings 
into optimization problems by introducing different notions of cost 
to measure deviation from the task's requisite frequency
\cite{SST09, IK18}. 

The frequency condition for PS
is that task $i$ must be scheduled
at least once in the interval $[m, m + a _i)$
starting on any day $m$.
We could consider a somewhat simpler problem \cite{LL73, BCPV96, BL98} where
instead we require this condition 
only when the starting day $m$ is
an integer multiple of $a _i$. 
We may also consider variants in which
(there can be days to which no task is assigned, and)
each task~$i$ must be scheduled \emph{precisely}
once (rather than \emph{at least} once) in $a _i$ days~\cite{WL83, HM82, Sun92}.

The \emph{packing}-style problem considered in
this paper (in which the tasks are
packed into $\Zset$) is complementary to
a \emph{covering}-style variant of the problem,
in which task $i$ may only be scheduled \emph{at most once}
in $a_i$ days \cite{KS20, KKK25}.
For the covering problem, there was
a conjecture similar to Theorem~\ref{theorem} (namely,
that any instance of integer-valued periods
with a density of at least $1.26\ldots$ suffices to
cover $\Zset$) \cite[Conjecture 18]{KS20}. 
After the preliminary announcements of our Theorem~\ref{theorem} \cite{Kaw24a, Kaw24} were made, 
this conjecture has also been confirmed, 
independently in two papers \cite{KK26, Mis26}, 
by partly applying our techniques. 

\subsection*{Organization of the paper}
In Section~\ref{section: fractional}, 
we generalize PS to non-integer periods, 
and note some basic properties of this general problem, 
which, as we pause to note, allow a simple proof of Theorem
\ref{theorem: two distinct}.
In Section~\ref{section: main theorem}, we reduce our main theorem (Theorem~\ref{theorem})
to checking the schedulability of finitely many instances. 
Section~\ref{section: implementation} explains some techniques for this exhaustive scheduling. 
In Section~\ref{section: bamboo}, we discuss how our bound for PS can be used 
to establish Theorem~\ref{theorem: bamboo}.

\section{Fractional Periods}
\label{section: fractional}

As noted above, an essential component of our argument
is to extend the allowed values of periods 
to real numbers. 
We do so by requiring that 
a task~$i$ with possibly non-integer period $a _i > 0$ 
must be performed at least $\lfloor r / a _i \rfloor$ times during any $r$-day interval.
The new frequency condition for task $i$ is thus 
\begin{quote}
for each $r \in \Nset$ and $m \in \Zset$,
the set $[m, m + r) \cap S _i$
has at least $\lfloor r / a _i \rfloor$ elements,
\end{quote}
or equivalently that 
\begin{quote}
for each $s \in \Nset$ and $m \in \Zset$,
the set $[m, m + s \cdot a _i) \cap S _i$
has at least $s$ elements.
\end{quote}
For integer-valued $a_i$ this reduces to the original condition,
but for non-integer $a_i$ it acquires new significance.
For example, a period $a _i = \frac 7 2$ requires
that task $i$ be performed at least once every $4$ days
\emph{and} at least twice every $7$ days; this could be achieved
if we schedule task $i$ every week on Monday and Thursday,
but not if we schedule it on Monday and Wednesday,
or if we simply schedule it every $4$ days. 

One advantage of this extension is that it does not affect the validity of
the basic properties of schedulability stated in the following lemma, 
which have been used, explicitly or implicitly, in previous studies as well (for integer periods).
We write $A \sqcup B$ for
the instance consisting simply of tasks in $A$ and tasks in $B$; 
for example,
$(6, 6) \sqcup (4, 4, 6) = (4, 4, 6, 6, 6)$.

\begin{lemma}
\label{lemma: basic properties}
If $A \sqcup (a)$ is schedulable, then so are 
\begin{itemize}
\item
$A \sqcup (b)$, for any period $b \geq a$
(weakening), and 
\item
$A \sqcup (\underbrace{a \cdot q, \dots, a \cdot q} _{q})$,
for any positive integer $q$
(splitting).
\end{itemize}
\end{lemma}

\begin{proof}
For the weakening property, 
simply schedule the period-$b$ task on the days reserved for the period-$a$ task.
For the splitting property,
schedule the $q$ new tasks, in a cyclically repeating sequence, 
on the days reserved for the period-$a$ task.
\end{proof}

This is the only preparation we need to give a concise proof
of Theorem \ref{theorem: two distinct}, which we pause to state here
(though it is not needed for Theorem \ref{theorem}).

\begin{proof}[Proof of Theorem \ref{theorem: two distinct}.]
By the splitting property of Lemma \ref{lemma: basic properties},
we may assume we have just one task with each period value,
i.e., our instance is
$(a _1, a _2)$ with $1 / a _1 + 1 / a _2 \leq 1$.
By the weakening property,
we may further assume 
$1 / a _1 + 1 / a _2 = 1$.

Consider a circle with circumference~$1$,
and two particles $i \in [2]$ moving along it in opposite directions, 
each with daily speed $1 / a _i$
(Figure~\ref{figure: circle}), 
\begin{figure}[t]
\centering
\includegraphics[scale=1.02]{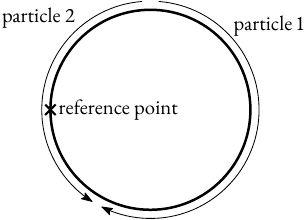}
\caption{Two particles moving in opposite directions, together covering the whole circle exactly each day.}
\label{figure: circle}
\end{figure}
so that they meet each other every midnight, 
which is possible because $1 / a _1 + 1 / a _2 = 1$.
Fix any point on the circle and call it the reference point.
During each day, 
the two particles together cover the circumference exactly once,
and thus exactly one of them sees the reference point,
under the convention (just in case the particles happen to meet at the reference point) that
midnight belongs to the next day for particle~$1$,
and to the previous day for particle~$2$.
This allows us to schedule each task $i \in [2]$
on days when particle~$i$ sees the reference point.
This schedule satisfies the frequency condition for both tasks~$i$,
because the time at which particle~$i$ is at the reference point
and the time at which it comes back there after going around the circle $s \in \Nset$ times 
are $s \cdot a _i$ apart,
and thus they belong to days that are at most $\lceil s \cdot a _i \rceil$ apart. 
\end{proof}

The fact that we obtain two complementary sets of integers in this way
is sometimes referred to as Rayleigh's or Beatty's theorem,
especially when $a _1$ and $a _2$ are irrational. 

\section{Proof of the Main Theorem}
\label{section: main theorem}

\newcommand{\fold}{\mathit{fold}}
\newcommand{\round}{\mathit{round}}

Let $\mathscr A$ be the set of instances 
in Theorem~\ref{theorem}, i.e., 
those with integer periods and density $\leq \frac 5 6$. 
This is an infinite set, 
and we want to prove all its instances schedulable
by reducing them to finitely many instances 
so that schedulability can be checked exhaustively. 
For this purpose, 
we repeatedly apply the procedures of Lemma~\ref{lemma: basic properties} \emph{in reverse} 
to each instance in $\mathscr A$, 
obtaining an instance
that has only small periods.
Consider the following process
which can be applied to any instance
whose largest period $a$ exceeds a threshold $\theta > 0$: 
\begin{itemize}
\item
if the instance contains only one task with period $a$, 
decrease its period to the second largest period $> \theta$, 
or to $\theta$ in case $a$ is the only period $> \theta$; 
\item 
if the instance contains multiple tasks with period $a$,
replace two of them with a single task with period $a / 2$.
\end{itemize}
Given an instance $A$ and $\theta > 0$,
every second application of this process
strictly decreases the number of tasks with period $> \theta$,
so we eventually arrive at an instance with all periods $\leq \theta$.
We write $\fold _\theta (A)$ for the resulting instance
(because Gąsieniec, Smith and Wild \cite[Section 5.2.2]{GSW22} call a variant of this process \emph{folding}). 
For example, $\fold _{22} (3, 4, 8, 17, \allowbreak 42, 55, 72) = (3, 4, 8, \frac{55}{4}, 17)$. 

\begin{lemma}
\label{lemma: folding}
For any PS instance $A$ and any $\theta>0$,
\begin{enumerate}
\item 
If $\fold _\theta (A)$ is schedulable, so is $A$.
\item
Any task in $\fold _\theta (A)$
with period $\leq \theta / 2$ is already in $A$. 
\item
$\density (\fold _\theta (A)) < \density (A) + 1 / \theta$, 
where the density of an instance with possibly non-integer periods
is still defined as the sum of the reciprocals of the periods.
\end{enumerate}
\end{lemma}

\begin{proof}
The two operations used in turning $A$ into $\fold _\theta (A)$ 
\begin{enumerate}
\item
are the inverses, respectively, of 
the weakening and splitting operations
in Lemma~\ref{lemma: basic properties}, 
which preserve schedulability, 
\item
only create periods that are $\geq \theta$ and $> \theta / 2$, respectively, 
\item
and never increase   the difference between 
the density and the reciprocal of the largest period. 
Note that this reciprocal remains $\leq 1 / \theta$ 
until the very last possible application of the second operation 
(which does not affect the density). 
\qedhere
\end{enumerate}
\end{proof}

Thus, 
given $A \in \mathscr A$, 
the resulting instance 
$\fold _\theta (A)$ 
belongs to the set $\mathscr B _\theta$ 
of instances that 
have periods in $((0, \theta / 2] \cap \Nset) \cup (\theta / 2, \theta]$
and density $\leq \frac 5 6 + 1 / \theta$. 
But $\mathscr B _\theta$ is still an infinite set. 
Given $B \in \mathscr B _\theta$, 
we define $\round _\theta (B)$ to be 
the instance obtained from $B$ by replacing every period $b > \theta / 2$ 
by $\lceil b \rceil - 1$, the largest integer less than $b$. 
For example, $\round _{22} (3, 4, 8, \frac{55}{4}, 17) = (3, 4, 8, 13, 16)$.
Again by the weakening property of Lemma~\ref{lemma: basic properties}, 
$B$ is schedulable if $\round _\theta (B)$ is. 
Note that $\density' (\round _\theta (B)) \leq \density (B)$, 
where
\begin{equation*}
  \density' \bigl( (c _i) _{i \in [k]} \bigr) =
  \sum _{i \in [k]}
  \begin{cases}
    \dfrac{1}{c _i} & \text{for $c _i < \biggl\lfloor \dfrac \theta 2 \biggr\rfloor$}, 
    \\[10pt]
    \dfrac{1}{c _i + 1} & \text{for $c _i \geq \biggl\lfloor \dfrac \theta 2 \biggr\rfloor$}.
  \end{cases}
\end{equation*}
We have thus reduced the schedulability of $A$ to
that of $\round _\theta (\fold _\theta (A))$, 
which belongs to the set $\mathscr C _\theta$ of 
instances $C$ that consist of integer periods $< \theta$ and 
satisfy $\density' (C) < \frac 5 6 + 1 / \theta$. 
Theorem~\ref{theorem} follows if we 
manage to schedule all instances in 
the finite set $\mathscr C _\theta$. 
Of course, there is a danger that 
our reduction has been too sloppy, 
so that $\round _\theta (\fold _\theta (A))$ may be unschedulable 
despite $A$ being schedulable. 
But in fact, choosing $\theta = 22$ proves this strategy successful: 
we have verified the following lemma (and hence Theorem~\ref{theorem})
by exhaustive computer search
as we explain in the next section. 

\begin{lemma}
  \label{lemma}
  Every instance in $\mathscr C _{22}$
  is schedulable.
\end{lemma}

\section{Scheduling Small Instances (Verifying Lemma~\ref{lemma})}
\label{section: implementation}

Although we have reduced Theorem~\ref{theorem} to 
Lemma~\ref{lemma}, which can be verified exhaustively in principle, 
we still face a substantial computational challenge. 
The previously best experimental result of this kind \cite{GSW22} 
has shown that every instance with $12$ tasks and density $\leq \frac 5 6$
is schedulable. 
We need a significant improvement over it, 
not only because our set $\mathscr C _{22}$ simply has far more instances, 
but also because these instances 
are typically longer and denser 
and hence are harder to schedule. 
In this section, 
we describe some techniques we used 
to carry out the computation in a reasonable amount of time. 
A computer program implementing these ideas
is publicly available on GitHub at
\begin{quote}
\url{https://github.com/a7kawamura/pinwheel_solver} .
\end{quote}

\subsection*{State Transition Graphs}
\label{subsection: deciding schedulability}

Although solutions to the PS problem are infinite schedules,
the following discussion shows \cite[Theorem 2.1]{HMRTV89} that
any schedulable instance $A = (a _i) _{i \in [k]}$
has a schedule that repeats a finite sequence of tasks
(perhaps justifying the name of ``pinwheel'' scheduling).

The elements of 
$[a _1] \times \dots \times [a _k]$
are called \emph{states};
if, at the end of a day, we are in a state $(u _i) _{i \in [k]}$,
this means that
each task $i$ must be performed no later than $u_i$ days from now. 
Performing task $j \in [k]$ on the next day
will bring us into state $u' = (u' _i) _{i \in [k]}$, where
\begin{equation*}
  u' _i
  =
\begin{cases}
  a _i & \text{if $i = j$}, \\
  u _i - 1 & \text{otherwise}.
\end{cases}
\end{equation*}
We write $u \vdash u'$
if this is the case for some $j$. 
An instance $A$ is thus schedulable if and only if
its \emph{state transition graph}, i.e.,
the directed graph with states as nodes and the relation $\mathord\vdash$ as edges,
admits an infinite walk,
or equivalently,
contains a cycle.
This is why any schedulable instance has a cyclic schedule, 
and also why schedulability can be checked in finite amount of time
by simply searching this graph for a cycle. 

The challenge is that 
the number of states in the graph
is exponential in the size of the input $A$,
and there indeed exist families of schedulable instances 
whose graphs have only exponentially long cycles.
Thus, although the problem of deciding schedulability 
is in $\classPSPACE$~\cite[Corollary 2.2]{HMRTV89} by Savitch's theorem,
we do not know whether it is in, say, $\classNP$.
There is evidence \cite{KL25} that it is not likely in $\classP$, 
and some even conjecture \cite{BvEMSRS22} that
it is $\classPSPACE$-complete. 

This lack of a notion of a solution of polynomial length (certificate of schedulability) 
gives rise to some subtlety in discussing algorithms that are supposed to ``output'' a schedule. 
One (somewhat informal) way to formulate efficient scheduling in this context is
to require that we can, given an instance, generate in polynomial time a \emph{fast online scheduler} (FOLS),
i.e., a program that efficiently computes which task to perform each day~\cite{HMRTV89}.
Many of the results about PS or BGT, including our Theorem~\ref{theorem: bamboo}, 
give such algorithms. 

\subsection*{Techniques for Fast Scheduling}
We thus have to search an extremely large graph, 
but there are some ways to reduce the computational cost 
by exploiting the way this graph is built. 

One useful observation is that 
if the state transition graph has a cycle,
then the cycle is reachable from the state $(a _i) _{i \in [k]}$. 
Thus, we can simply start a depth-first search 
at this initial state, 
while maintaining the path to the current state until this path makes a cycle. 
We also keep the set of 
dead states found so far, 
i.e., those that never lead to a cycle: 
for any $i \in \Nset$, a state with more than $i$ entries $\leq i$ is dead
(this is a benefit of our ``urgency'' formulation of states \cite[Section~4.3]{GSW22}), 
and a state that only leads to dead states is dead. 

Another improvement comes from the observation that, 
since our instances have periods up to $21$, 
and the hard instances are those with $12$--$18$ tasks, 
they typically have multiple tasks with the same period. 
We thus identify states up to permuting them. 
This reduces not just the number of states, 
but also the out-degree of each state,
because we may assume that 
among multiple tasks with the same period, 
we always perform the least recently performed one. 
This significantly reduces the number of states and edges, 
and also often the length of the sought repeating pattern. 
For example, 
instead of saying that the instance $(2, 4, 8, 8)$ 
has a schedule that repeats the $8$-day pattern $(1, 2, 1, 3, 1, 2, 1, 4)$ of tasks, 
we can more economically say that 
this schedule just repeats the $4$-day sequence $(2, 4, 2, 8)$ of periods, 
with the understanding that the two period-$8$ tasks always take turns. 

\subsection*{Reducing the Number of Instances to Check}
Given that scheduling each individual instance can take a significant amount of time, 
it is important to cut down on computation 
so that we do not have to check 
all the $25\,592\,972$ instances in $\mathscr C _{22}$. 

A trivial but useful idea to do so 
is that, by the weakening property of Lemma~\ref{lemma: basic properties}, 
we can skip
instances that component\-wise dominate 
another instance in $\mathscr C _{22}$.
In our program, we call the remaining $676\,225$ instances \emph{critical}. 

We may also hope to reduce the number of instances 
by using a value smaller than $22$ in Lemma~\ref{lemma}. 
Unfortunately, we cannot: for example, 
the instance $(3, 4, 7, 10, 15) \in \mathscr C _{20}$ is not schedulable,
even though $\frac 1 3 + \frac 1 4 + \frac 1 7 + \frac{1}{11} + \frac{1}{16} < \frac 5 6 + \frac{1}{20}$
(it is still possible, for all we know, that all instances in $\mathscr B _{20}$ are schedulable). 
But such instances are not so common: 
most instances in $\mathscr C _{22}$
remain schedulable even if we ``apply folding a bit further''. 
For an instance $C = (c _i) _{i \in [k]}$ of length $k > 1$, 
define $C' = (c _i) _{i \in [k - 2]} \sqcup (\lfloor c _{k - 1} / 2 \rfloor)$ as 
the result of applying to $C$
a single step of folding and then rounding down. 
Again by Lemma~\ref{lemma: basic properties}, 
$C$ is schedulable if $C'$ is.
Since $C'$ has fewer tasks than $C$, 
it is often easier to schedule, 
although of course there is a risk that $C'$ may not be schedulable while $C$ is.
Thus, to check the schedulability of $C$, 
we run the cycle-detecting algorithm above 
on instances $C$, $C'$, $C''$, \ldots\ (some of which may not be in $\mathscr C _{22}$) in parallel, 
until one of them turns out schedulable. 

This idea of continued folding causes many instances 
to be proved schedulable via a common short instance, 
thereby reducing the number of instances that 
we need to actually schedule. 
We ended up completing the cycle-detecting algorithm
only about $60\,000$ times
(this number fluctuates 
depending on which of the parallel searches finish first).

\section{Approximation Algorithm for BGT}
\label{section: bamboo}

We will provide a polynomial-time algorithm $M$ that,
given an integer PS instance $(a _i) _{i \in [k]}$, either accepts or rejects
so that
\begin{itemize}
\item
if $M$ rejects, 
$(a _i) _{i \in [k]}$ is not schedulable, and
\item 
if $M$ accepts, 
$\bigl( \bigl\lfloor \frac 4 3 \cdot a _i \bigr\rfloor \bigr) _{i \in [k]}$
is schedulable; 
\end{itemize}
note that for some instances, both answers are allowed. 
This implies Theorem~\ref{theorem: bamboo}: 
given a BGT instance $(h _i) _{i \in [k]}$,
we can binary search for an integer height~$G$ such that 
$M$ accepts $(\lfloor G / h _i \rfloor) _{i \in [k]}$ 
but not if $G$ is replaced by $G - 1$. 

Such an algorithm $M$ comes easily
if all periods $a _i$ are $> 2$: 
simply accept $(a _i) _{i \in [k]}$ if and only if it has density $\leq 1$. 
When $M$ accepts, 
the instance $\bigl( \bigl\lfloor \frac 4 3 \cdot a _i \bigr\rfloor \bigr) _{i \in [k]}$ 
has density $\leq \frac 5 6$ (unless $k = 1$)
and is hence schedulable by Theorem~\ref{theorem}. 
This is because elementwise we have 
$\bigl\lfloor \frac 4 3 \cdot a \bigr\rfloor \geq \frac 6 5 \cdot a$, 
as can be verified by 
\begin{equation*}
  \biggl\lfloor \frac 4 3 \cdot a \biggr\rfloor \geq \frac 4 3 \cdot a - \frac 2 3 \geq \frac 4 3 \cdot a - \frac{2}{15} \cdot a = \frac 6 5 \cdot a  
\end{equation*}
for $a \geq 5$ and individually for $a = 3$, $4$. 

Since this inequality does not hold for $a = 2$, 
We need special treatment for period $2$. 
But a period-$2$ task is actually easy to deal with: 
we can safely schedule it regularly every other day, 
since performing the same task for two days in a row would only make scheduling other tasks harder. 
Precisely, 

\begin{lemma}
\label{lemma: period 2}
A nonempty PS instance $(a _i) _{i \in [k]}$ with $a _1 = 2$ is schedulable 
if and only if 
the instance $\bigl( \bigl\lfloor \frac 1 2 \cdot a _{i + 1} \bigr\rfloor \bigr) _{i \in [k - 1]}$ is.
\end{lemma}

\begin{proof}
A schedule for the former can be obtained from one for the latter
by inserting a day for the new period-$2$ task every other day. 
A schedule for the latter can be obtained from one for the former
by removing all days for the period-$2$ task and closing up. 
\end{proof}

Using this fact, our algorithm~$M$ eliminates (recursively if needed) period $2$  
before scheduling other tasks. 
That is, given a PS instance $A = (a _i) _{i \in [k]}$: 
\begin{itemize}
\item 
If $A$ is empty, accept. 
\item 
If $a _1 = 2$,
recursively apply $M$ itself to the instance
$\bigl( \bigl\lfloor \frac 1 2 \cdot a _{i + 1} \bigr\rfloor \bigr) _{i \in [k - 1]}$.
\item 
If $a _1 \neq 2$,
accept if $A$ has density $\leq 1$, and reject otherwise.
\end{itemize}

Note that the second case is correct by Lemma~\ref{lemma: period 2}: 
the rejecting case is the ``only if'' direction of the lemma, 
and the accepting case is because 
the schedulability of 
the instance $\bigl( \bigl\lfloor \frac 4 3 \cdot \bigl\lfloor \frac 1 2 \cdot a _{i + 1} \bigr\rfloor \bigr\rfloor \bigr) _{i \in [k - 1]}$ implies 
that of the elementwise dominating instance 
$\bigl( \bigl\lfloor \frac 1 2 \cdot \bigl\lfloor \frac 4 3 \cdot a _{i + 1} \bigr\rfloor \bigr\rfloor \bigr) _{i \in [k - 1]}$, 
and hence of $\bigl( \bigl\lfloor \frac 4 3 \cdot a _i \bigr\rfloor \bigr) _{i \in [k]}$ 
by the ``if'' direction of the lemma. 
The third case is also correct: 
it is trivially so if $a _1 = 1$; 
otherwise, all periods are $> 2$, and thus the criterion of $\density (A) \leq 1$ works as discussed above. 

\section{Conclusion}

Our Theorem~\ref{theorem} thus enters the list of 
simply stated mathematical conjectures
that have been confirmed by a computer experiment  
while defying attempts at human-readable proofs for decades.
This list includes classics such as the 
Four-Colour Theorem \cite{RSST97} and Kepler Conjecture \cite{Hal05}, 
as well as more recent examples in enumerative combinatorics
reported in this journal \cite{KKZ09, KKZ11}. 
As the authors of both these last papers remark in their final paragraphs, 
we are still left with the intriguing human challenge of 
finding a short and elegant ``proof from The Book'', 
although they both point out the possibility that such a short explanation may not exist. 

Let us end with a conjecture 
that came out of the computer experiment for Theorem~\ref{theorem}
but which seems to require a better understanding of its proof. 
In the course of our experiment, 
it turned out that the bound $\frac 5 6$ was supported by very few instances: 
there is a positive constant $\varepsilon$ such that 
the only unschedulable instances with density $< \frac 5 6 + \varepsilon$
are of the form $(2, 3, a _3)$ or $(3, 4, 4, a _4)$. 
Instances of the form
$(a _1, a _1 + 1, \dots, \allowbreak a _1 + 1, a _{a _1 + 1})$ are likewise unschedulable, 
but have higher density for $a _1 > 3$. 
We suspect 
that these are the sparsest unschedulable instances starting with each fixed value of $a _1$: 

\begin{conjecture}
An integer PS instance $A = (a _i) _{i \in [k]}$ is schedulable if
$\density (A) \leq 1 - (a _1 - 1) / (a _1 (a _1 + 1))$.
\end{conjecture}

Since the results in this paper were announced \cite{Kaw24}, 
some authors have studied 
such relations between the density bound and the smallest period $a _1$ \cite{GJKLLMR24, KKK25, MRK26}. 

\subsection*{Acknowledgements}
The author is grateful to
Hiroshi Fujiwara, Leszek G\k{a}sieniec, Yusuke Kobayashi, Igor Potapov, Benjamin Smith and Sebastian Wild, 
as well as the anonymous referees, 
for valuable comments on earlier versions of this work.
He also thanks Keita Hiroshima for helping to verify the list of schedules in Lemma~\ref{lemma}.
This research was supported by JSPS KAKENHI (Grant-in-Aid for Scientific Research) JP23K28036 and by the Kyoto University Foundation.

\end{document}